\documentclass[aps,prx,twocolumn,secnumarabic,superscriptaddress,nofootinbib]{revtex4-2} 

\usepackage[english]{babel}
\usepackage[utf8]{inputenc}
\usepackage{longtable}
\usepackage{morefloats}
\usepackage[dvips]{graphicx,color}
\usepackage[dvipsnames]{xcolor}
\usepackage{epsfig,amsfonts,amsbsy}
\usepackage{amsmath,amsthm,amssymb}
\usepackage{bbold}
\usepackage{url}
\usepackage{verbatim}
\usepackage{array}
\usepackage{multirow}
\usepackage{tabularx}
\usepackage{braket} 
\usepackage{dsfont}
\usepackage{bm}
\usepackage{physics}
\usepackage[colorlinks=true, pdfstartview=FitV, linkcolor=blue,
citecolor=blue, urlcolor=blue]{hyperref}
\usepackage[capitalise]{cleveref}
\usepackage{titlesec}
\usepackage[ruled,longend]{algorithm2e}
\usepackage{hyperref}

\usepackage{textcomp}

\newcommand{\CXtime}{t_\text{CX}}
\newcommand{\MRtime}{t_\text{M/R}}

\begin{document}

\title{
Constant-depth magic state cultivation with Clifford measurements by gauging
}
\author{Bence Het\'enyi}
\email{bence.hetenyi@ibm.com}
\affiliation{IBM Quantum, IBM T. J. Watson Research Center, Yorktown Heights, NY 10598, USA} 
\author{Benjamin J. Brown}
\affiliation{IBM Quantum, IBM T. J. Watson Research Center, Yorktown Heights, NY 10598, USA} 
\affiliation{IBM Denmark, Sundkrogsgade 11, 2100 Copenhagen, Denmark}
\author{Dominic J.~Williamson}
\thanks{Current address: School of Physics, University of Sydney, Sydney, NSW 2006, Australia}
\affiliation{IBM Quantum, IBM Almaden Research Center, San Jose, CA 95120, USA}

\date{March 2026}

\begin{abstract}
Magic states are a scarce resource for two-dimensional qubit stabilizer codes. Magic state cultivation was recently proposed to reduce the cost of magic state preparation by measuring the transversal Clifford operator of the color code. 
Cultivation achieves $\sim 10^{-9}$ logical error rates for the $d=5$ color code, with substantially lower space-time overhead than magic state distillation. However, due to the $\mathcal{O}(d)$ depth of the Clifford measurement circuit, magic state cultivation becomes impractical for $d>5$. Here, we perform logical $XS^\dagger$ measurements on the color code by gauging a transversal Clifford gate, resulting in a constant-depth logical measurement circuit. We employ repeated gauging measurements with post-selection rather than performing error correction on the Clifford stabilizer code that emerges during the gauging protocol, thus gaining simplicity at the cost of scalability.  Our protocol requires a regular square grid connectivity and yields logical error rates comparable to magic state cultivation. The $d=7$ version of our protocol gives access to the $10^{-12}$ logical error rate regime at $0.05\%$ physical error rate while retaining more than $1\%$ of the shots after the equivalent of the cultivation stage. 
\end{abstract}

\maketitle

{\it Introduction.} Magic states are an essential resource for fault-tolerant quantum computation because they complement Clifford operations to form a universal gate set. Preparing magic states fault-tolerantly using two-dimensional codes is challenging due to the Bravyi-K\"onig bound which restricts relevant local transversal gates to be Clifford~\cite{bravyiClassificationTopologicallyProtected2013}. Recent work has revealed new approaches to overcome this limitation in two dimensional codes \cite{davydovaUniversalFaultTolerant,bauerPlanarFaulttolerantCircuits2025,huang2025generatinglogicalmagicstates,kobayashi2025cliffordhierarchystabilizercodes,huang2026hybridlatticesurgerynonclifford,warman2026transversalcliffordhierarchygatesnonabelian}, most approaches continue to rely on probabilistic preparation of magic states even in architectures with nonlocal connectivity \cite{yoderTourGrossModular2025,webster2026pinnaclearchitecturereducingcost}.

The most commonly used probabilistic approach is magic state distillation \cite{bravyiUniversalQuantumComputation2005}. Distillation is most commonly implemented via a logical circuit that consumes a number of noisy magic states to produce fewer magic states that are less noisy. The distillation circuit is post-selected on undesirable measurement outcomes, and so succeeds only probabilistically. Many improvements have been made to reduce the cost of magic state distillation, however, there is still a considerable overhead required to prepare logical magic states in the surface code compared to other logical gates \cite{bravyiUniversalQuantumComputation2005,fowlerSurfaceCodesPractical2012,litinskiMagicStateDistillation2019}.

Measuring a transversal Clifford operator of a self-dual code \cite{Goto2016Min,chamberlandFaulttolerantMagicState2019,Chamberland2020,Gupta2024,Itogawa2025} provides a simple alternative method to improve the fidelity of a non-stabilizer magic state, as opposed to distillation where the purification of the state requires multiple logical qubits and operations. A similar scheme that introduced several further optimizations, dubbed cultivation, was introduced in Ref.~\cite{gidneyMagicStateCultivation2024} where the logical $XS^\dagger$ operator of the color code was measured using a circuit of depth $O(d)$ while respecting square grid connectivity. Cultivation achieved a logical error rate of $6\cdot 10^{-10}$ in the $d=5$ color code with a success probability of $15\%$ at $0.1\%$ physical error rate. A number of follow up works on Cultivation have appeared more recently~\cite{vaknin2026efficientmagicstatecultivation,chen2025efficientmagicstatecultivation,sahay2025foldtransversalsurfacecodecultivation,claes2025cultivatingtstatessurface}. 

Gauging provides an alternative approach to implementing a logical Clifford measurement with a low-depth adaptive quantum circuit~\cite{williamsonLowoverheadFaulttolerantQuantum2024,davydovaUniversalFaultTolerant}. 
These circuits were introduced in the context of creating topological states of matter~\cite{Williamson2020a,Tantivasadakarn2021,Tantivasadakarn2022}. 
Implementing the logical measurement via gauging has the highly desirable property that it reduces the circuit depth of the Clifford measurement step to $O(1)$.
While the scheme of Ref.~\cite{williamsonLowoverheadFaulttolerantQuantum2024} is applicable to Clifford operators, constructing a suitable ancilla system that respects the geometric constraints presents a challenge.

Here, we provide a practical implementation of the gauging logical Clifford measurement protocol using a planar network of ancilla and flag qubits. Our scheme has the same hardware requirements as Ref.~\cite{gidneyMagicStateCultivation2024}, but performs the Clifford measurement step in constant circuit depth. We show that the $d=7$ version of our protocol can achieve logical error rates below $5\cdot10^{-12}$ for $0.05\%$ physical error rate while retaining $1.5\%$ of the shots after post selection.

\begin{figure*}
    \centering
    \includegraphics[width=.8\linewidth]{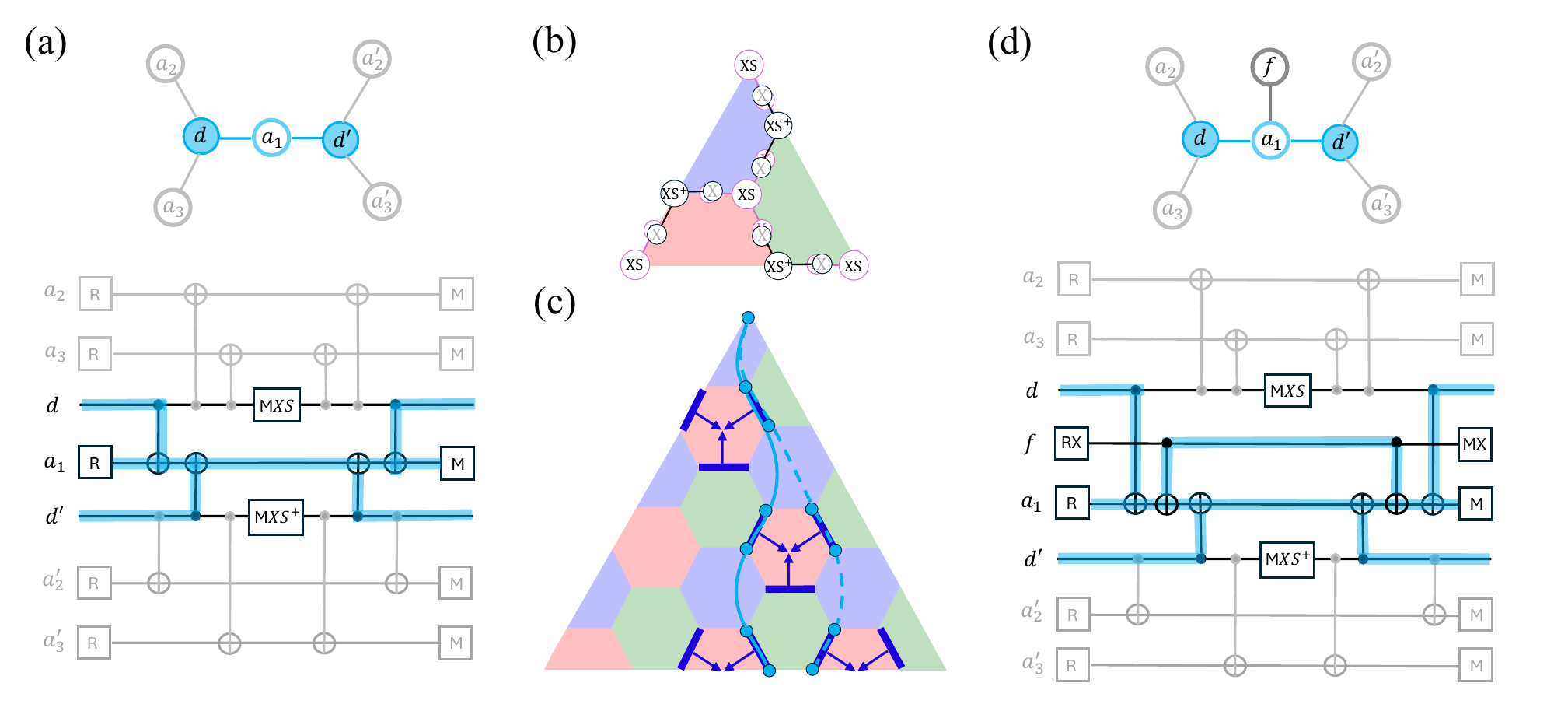}
    \caption{(a) Gauging measurement of the logical $XS^\dagger$ operator supported on data qubits ($d$ and $d'$) using ancillas ($a$ and $a'$) on the edges of the honeycomb lattice. The weight of logical Z operator (cyan) is reduced during the measurement. (b) Gauge operators for measuring the transversal $XS^\dagger$ on the $d=3$ color code. (c) Two minimum-weight logical representatives (solid and dashed cyan lines) of the $d=7$ color code. Arrows point at the flag qubit shared by the same colored edges. (d) Flagged version of the circuit in (a), where an extra qubit ($f$) increases the weight of the logical before the the circuit distance would drop.}
    \label{fig:flagged_circuit}
\end{figure*}

{\it Gauging Clifford measurement.} In this section, we briefly summarize the gauging logical measurement scheme introduced in Ref.~\cite{williamsonLowoverheadFaulttolerantQuantum2024}. Consider a transversal logical operator of the form $C_L=\prod_i U_i$, where $U_i = U_i^{-1}$ and $i$ runs over the data qubits. Note, $U_i = U_i^{-1}$ holds for all Hermitian Clifford operators. We define an ancilla system where each ancilla is connected to a pair of data qubits and is prepared in the $\ket 0$ state at the beginning of the protocol. Gauss-law operators are defined for every data qubit as follows
\begin{equation}
A_i = U_i\prod_{a\in \partial i}X_a. 
\end{equation}
The product of these operators is the logical operator $C_L$ we aim to measure. In Ref.~\cite{williamsonLowoverheadFaulttolerantQuantum2024} it is shown that measuring these gauge operators satisfies {\it (i)} the product of measurements $\sigma_L=\prod_im_i$ is the logical eigenvalue and {\it (ii)} the measurement outcome of the ancilla qubits is sufficient to restore the correct post-measurement state $(1+\sigma_L C_L)\ket{\psi}_Q$ by applying the correction operator $\prod_{i\in Q'}U_i$ on a subset of data qubits $Q'\subset Q$ (see SM Sec.~\ref{sec:gauging_proof} for the proof).

We define the ancilla graph $G$ by identifying data qubits with vertices and ancilla qubits with the edges of the graph. In order to achieve fault tolerance and constant qubit overhead, it is sufficient for the ancilla graph $G$ to fulfill the following properties {\it (i)} the length of the edge path between any pair of qubits included in a stabilizer should be constant, {\it (ii)} the Cheeger constant of the graph $h(G)$ should be sufficiently large, and {\it (iii)} the graph $G$ should admit a sparse cycle basis~\cite{williamsonLowoverheadFaulttolerantQuantum2024}. 

In the case when $U_i$ are Hermitian Clifford operators that are not Pauli, there is an additional criterion that needs to be fulfilled, independent of how the Clifford operator is measured. In a self-dual CSS code, at least one type of stabilizers, $X$ or $Z$, will not generally commute with the transversal Clifford operator. Commutation is only achieved when the state is in the mutual $+1$ eigenspace of all the stabilizers of the other type.

\begin{figure*}[t]
    \centering
    \includegraphics[width=1.
    \linewidth]{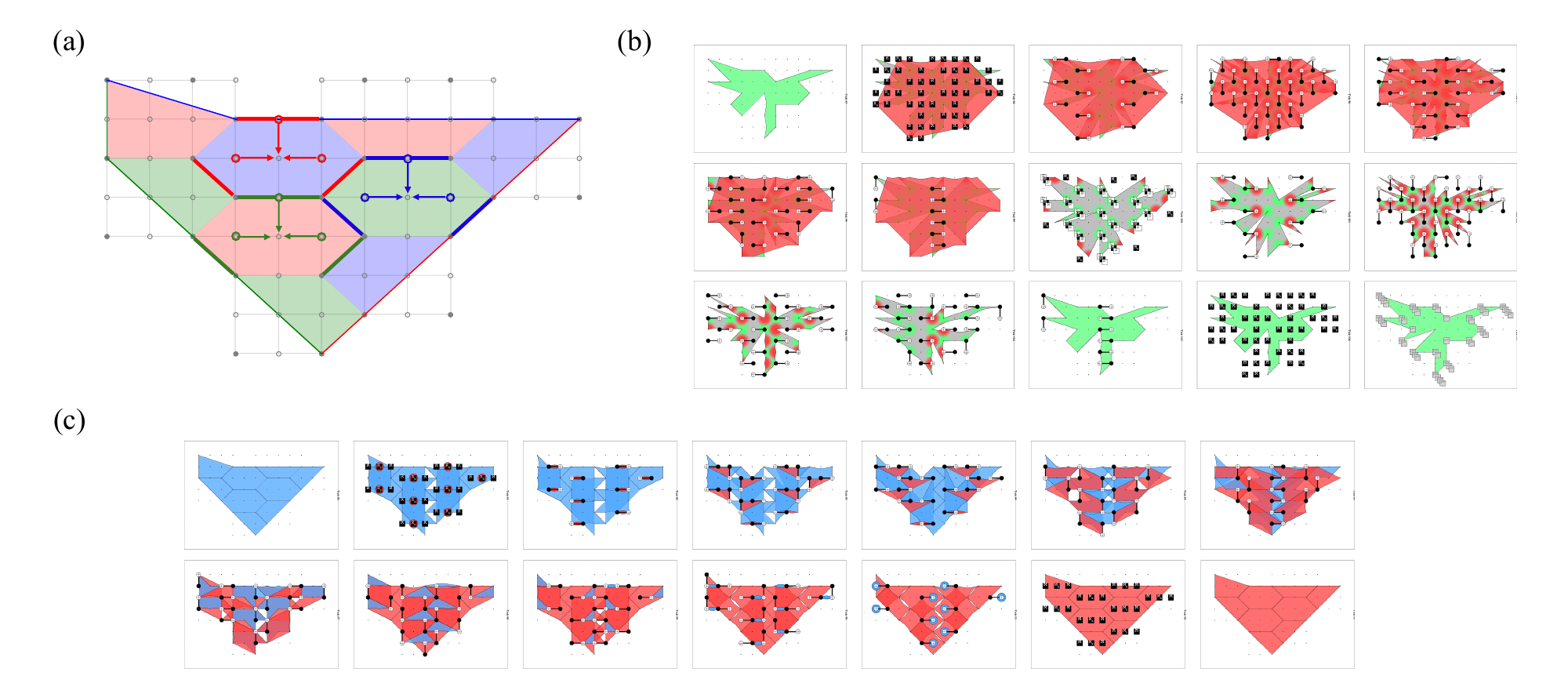}
    \caption{(a) Qubit layout for the square-grid implementation of the gauging Clifford measurement. In the middle of each plaquette there is a flag qubit, that flags the three edges that are parallel with one of the boundaries. The top left corner is deformed to reduce the number of extra qubits (needed for the ancilla cycles) at the boundary. (b) Circuit schedule of the a gauging logical (Y) measurement step. The two detector slices show a contracting (exists in steps 1 to 8) and an expanding logical Y check (steps 2 to 15). (c) Circuit schedule for one stabilizer measurement round. Detector slices correspond to the contracting Z (blue) and the expanding X detectors (red). Contracting X and expanding Z detectors are not shown on the figure for clarity.}
    \label{fig:layout_detslices}
\end{figure*}

{\it Gauging logical Clifford measurement on the color code.} We now consider the honeycomb color code with triangular boundary conditions~\cite{Bombin2006}, see SM Sec.~\ref{sec:colorcode_hhx} for a review. This code is self dual and has weight-six (weight-four) X and Z plaquette stabilizers in the bulk (at the boundary). Logical Pauli, as well as Clifford operators are transversal (e.g., $XS^\dagger$). The plaquettes are three-colorable and we assign the color of the weight-four plaquette at each corner of the triangle to the respective corner and the edge opposite to it. Low-weight logical Pauli operators can be constructed by finding a Pauli string between a corner and the corresponding edge such that it goes through plaquettes of the same color (see Fig.~\ref{fig:flagged_circuit}(c)).

To measure the transversal logical $XS^\dagger$ operator of the color code we use an ancilla system such that every pair of adjacent vertices on the honeycomb lattice is connected to the same ancilla qubit~\cite{williamsonDynamicalQuantumCodes2025}. This way, the data and ancilla qubits form a heavy hexagonal lattice (see Fig.~\ref{fig:flagged_circuit}(b) and SM Sec.~\ref{sec:magicstate_hhx}). The resulting ancilla graph $G$ is connected, contains short loops but has a Cheeger constant $h(G)<1$. In this case the distance of the protocol may be reduced (see Sec.~\ref{sec:cheeger}).

Gauge operators are measured by contracting their X support onto the respective data qubits, which are subsequently measured in the basis of their Clifford support. This scheme has a clear bottleneck, independent of the Cheeger-constant of the ancilla graph. Fig. \ref{fig:flagged_circuit}(a) shows how a $ZZ$ operator on two data qubits is contracted to a single ancilla qubit during the measurement of the gauge operators. Since this weight-reduction happens along every edge of the lattice, the minimum-weight logical representatives shrink to $\lceil d/2\rceil$ qubits.

{\it Gauging Clifford measurement with flag qubits.} In order to avoid the weight reduction of a logical $Z$ operator shown in Fig.~\ref{fig:flagged_circuit}(a), we employ a flag qubit $f$, adjacent to the ancilla $a_1$, prepared in the $\ket{+}$ state. Applying a CX gate, controlled on the flag and targeted on the ancilla, before the step where the distance-reduction happens, the logical Z operator expands to the flag qubit. This way the fault distance is maintained throughout the protocol. This is shown in Fig.~\ref{fig:flagged_circuit}(d).

Next, we argue that one flag qubit per plaquette is sufficient to avoid the weight reduction of every minimum-weight logical Z representative. As shown in Fig.~\ref{fig:flagged_circuit}(c), minimum-weight logicals that connect the blue corner to the blue boundary of the color code are supported on edges that connect blue plaquettes. None of these representatives can contain two of such edges around the same plaquette. Therefore, one may use the same flag qubit, to avoid the distance reduction along all three edges of a given plaquette. This leads to a factor of three reduction of flag qubit count.

In order to find a suitable two-dimensional qubit layout for the flagged protocol, we need a lattice that {\it (i)} supports data qubits in a hexagonal sublattice, {\it (ii)} has one ancilla qubit in-between every nearest-neighbor pair of data qubits, and {\it (iii)} has one flag qubit per plaquette, connected to three ancilla qubits around it. The embedding of such a lattice in a regular square grid is shown in Fig.~\ref{fig:layout_detslices}(a). There are additional data qubits and ancillas around the boundary of the lattice. These ensure that there are complete cycles in the ancilla graph even at the boundary plaquettes as required for the ancilla system in the third condition of the gauging protocol \cite{williamsonLowoverheadFaulttolerantQuantum2024}.

Having all the necessary components for the gauging logical measurement in the coupling graph, we turn to the syndrome measurement circuit of the color code. To this end we use the superdense syndrome measurement circuit of Ref.~\cite{gidney_new_2023} and prepare the ancilla Bell pair on the two ancillas that are inside each plaquette through the flag qubits. Since every data qubit on a given plaquette is the nearest neighbor of one of the ancillas, we can follow Ref.~\cite{gidney_new_2023} to encode the X and Z plaquette information into the Bell state of the ancillas. Through a series of circuit simplification steps (detailed in SM Sec.~\ref{sec:opt}) we can reduce the circuit depth of the stabilizer measurement to 10 CX layers. See Fig.~\ref{fig:layout_detslices}(c) for the contracting Z-type (expanding X-type) detecting regions.

\begin{figure*}[t]
    \centering
\includegraphics[width=1.\linewidth]{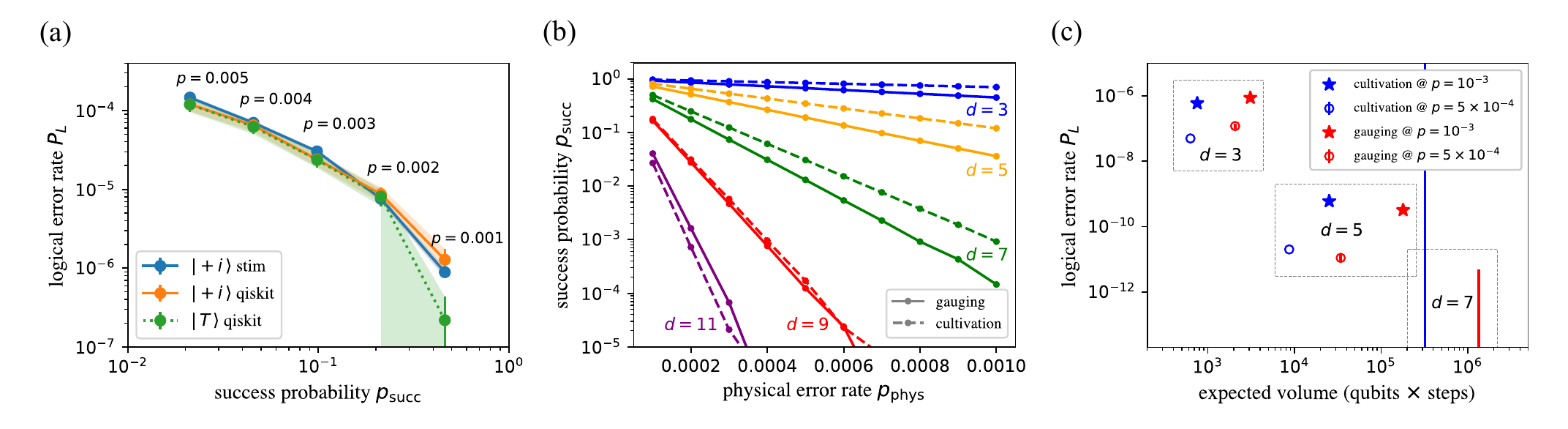}
    \caption{(a) Comparison of different simulation methods for the $d=3$ gauging logical measurement protocol under uniform circuit-level noise for various physical error rates $p$. The Clifford simulations of the $\ket{+i}$ state preparation, obtained via stim, align with that of the statevector simulation and the $\ket{T}$-state preparation. (b) Comparison between the gauging Clifford measurement and magic state cultivation for uniform circuit-level noise. For this noise model the break-even point in success probability is $d=9$. (c) Comparison of the logical error rate and the expected circuit volume. For $d=7$ there is no data available for cultivation, and we have not found a single logical failures in $2\times 10^{11}$ samples of the gauging protocol.
    }
    \label{fig:simulations}
\end{figure*}

The stabilizer measurements in our scheme are always followed by a gauging Clifford measurement step (see Sec.~\ref{sec:opt}),
which gives further opportunities for pipelining the circuit. This allows us to perform one syndrome and one Clifford measurement using 18 CX and 5 measurement (reset) layers (see Fig.~\ref{fig:pipelining} for details). For context, cultivation achieves an equivalent task in ${\frac 3 2 (d+5)}$ CX and 4 measurement (reset) layers.

To prepare magic states with larger fault distance $d_f$ we start from the $d=3$ example and prepare magic state non fault-tolerantly using a unitary circuit. This is followed by a stabilizer measurement and a gauging Clifford check at $d_\text{int}=3$. We grow the lattice by preparing Bell states on the edges that match the color of the boundary we are growing along. We continue this until we reach the final target distance, where we perform an additional $(d+1)/2$ rounds of stabilizer and Clifford check pairs (see SM Sec.~\ref{sec:growing}).

{\it Simulations and comparison with cultivation.} In order to verify that the gauging Clifford measurement scheme prepares the desired state, we performed full statevector simulations using \texttt{qiskit} on the $d=3$ example (25 qubits in total) \cite{aleksandrowicz_qiskit_2019}. The simulated protocol consists of a unitary $\ket T$-state preparation and two rounds of stabilizer and Clifford checks for uniform circuit level noise with physical error rates between $0.1\%$ and $0.5\%$ (see Fig.~\ref{fig:simulations}(a)). The uniform circuit-level noise model considered here is equivalent to that of Ref.~\cite{gidneyMagicStateCultivation2024} and we used the pipelined circuit in the simulation. We observe good agreement between the statevector simulations and the $\ket{+i}$ version simulated with \texttt{stim} \cite{gidney_stim_2021}. In both cases, the logical error probability is obtained by Monte-Carlo sampling and full post-selection. For the \texttt{qiskit} simulations we project the final state onto a stabilizer state and evaluate the expectation value of the 
$XS^\dagger$ operator on a shot-by-shot basis, while for the \texttt{stim} simulations we perform a noiseless stabilizer and final logical measurement in the Y basis. In both cases, the post selection on the final perfect round is included in the success probability.

\begin{table}[b]
    \centering
    \begin{tabular}{cccc}
        \hline
        $p_\text{phys}$ & $d_f$ & $P_L$ & $p_\text{succ}$ \\        
        \hline
        $0.1\%$ & 3 & $(8.8\pm 0.8) \times 10^{-7}$ & $46\%$\\
         & 5 & $(3.2\pm 0.4) \times 10^{-10}$ & $4\%$\\
         & 7 & $-$ & $0.01\%$\\
        \hline
        $0.05\%$ & 3 & $(1.2\pm 0.2)\times 10^{-7}$ & 65\%\\
         & 5 & $(1.1\pm 0.3)\times 10^{-11}$ & $20\%$\\
         & 7 & $<5\times 10^{-12}$ & $1.5\%$\\
        \hline
    \end{tabular}
    \caption{Logical error probability $P_L$ and success rate $p_\text{succ}$ of the $\ket{+i}$ state preparation for different fault-distances $d_f$ and physical error rates $p_\text{phys}$. For $d_f=7$ at $p_\text{phys} = 0.1\%$ we did not perform Monte-Carlo sampling due to the small success rates, and for $p_\text{phys} = 0.05\%$ we did not observe any logical failure in $2\times10^{11}$ post-selected samples. }
    \label{tab:simulation}
\end{table}

In Tab.~\ref{tab:simulation} we show the relevant figures of merit at higher fault distances. While the success probability is somewhat lower than that of cultivation, the logical error rates are comparable for $d_f\in\{3,5\}$ \cite{gidneyMagicStateCultivation2024}. We also benchmark our protocol for $d=7$ at $p_\text{phys} = 0.05\%$ and find an acceptable success probability for preparing a logical magic state with $\sim10^{-12}$ logical error rate. We note that a comparison for higher distances and lower error rates would be highly interesting due to the asymptotic advantage of the gauging Clifford measurement scheme in circuit depth. Unfortunately, we do not have data for cultivation at $d>5$.

Fig.~\ref{fig:simulations}(b) shows the success probability of the gauging Clifford measurement vs. cultivation. 
Under the uniform circuit-level noise model used in both works, the breakeven point in success probability only comes as $d=9$. This is due to the qubit overhead of the gauging protocol (see SM Sec.~\ref{sec:comp_cult} for detailed comparison). The latter uses $50\%$ more qubits, resulting in similar number of fault locations only at $d=9$. 

Arguably, the more important metrics are the expected circuit volume and the logical error rate. Fig.~\ref{fig:simulations}(c) shows such a comparison for smaller fault distances. At $d=3$ cultivation has both lower logical error rate and expected volume, the gauging protocol outperforms Gidney {\it et al.} in logical error rate already at $d=5$ and continues to improve the logical error rate for $d=7$. While $d=7$ data is not available for cultivation, and only an upper bound can be provided for gauging ($P_L<5\times10^{-12}$), we believe that our $d=7$ protocol can be a compelling alternative in the low error rate regime even if the expected circuit volume is higher than that of cultivation.

{\it Discussion.} In this work we described a method to prepare logical magic states on a square grid by implementing post-selected Clifford measurement via gauging. 
Our scheme has the advantage of maintaining a constant circuit depth for the Clifford measurement. 
There are some additional advantages compared to the cultivation, by dividing the lattice into two sublattices in a checkerboard-like pattern, data and flag qubits are occupying the same sublattice and ancillas are hosted on the other one. This separation reduces cross-talk between data qubits and has further advantages in architectures where data and ancilla qubits are distinguished at a physical level \cite{lacroixScalingLogicColor2024}.

Although direct comparison of the probability of success did not produce a clear advantage for $d=7$, we note that there are further opportunities for optimization both in the circuit depth and the count of qubits. The space time overhead of the protocol could be optimized using a time-dynamic approach instead of designating the roles of data, ancilla, and flag to different qubits \cite{mcewenRelaxingHardwareRequirements2023,shawLoweringConnectivityRequirements2025a}.

Our work points to an interesting future direction of numerically and experimentally benchmarking a scalable error-corrected protocol for magic state preparation via gauging Clifford measurement. This requires the error-correction with a Clifford stabilizer code and a just-in-time decoder as explained in Ref.~\cite{davydovaUniversalFaultTolerant}. While initial results on implementing the measurement of relevant Clifford stabilizers on a planar geometry are presented in SM Sec.~\ref{sec:magicstate_hhx}, further work to improve the fault-tolerance of these circuits would be advantageous. Finally, it would be interesting to understand the perfomance landscape of protocols that interpolate between post-selected error detection and full error correction for the preparation of magic states via logical Clifford measurement at a target error rate and footprint.

\section*{Acknowledgements}

We thank Tomas Jochym-O'Connor for insightful discussions. The scripts used to generate the data for this work are available at \href{https://github.com/hetenyib/gauging_clifford_measurement}{https://github.com/hetenyib/gauging\_clifford\_measurement}.

\bibliography{references}

\clearpage


\onecolumngrid
\pagebreak

\clearpage
\onecolumngrid
\vspace*{0.1cm}
\begin{center}
	\large{\bf Supplemental Material to \\`Constant-depth magic state cultivation with Clifford measurements by gauging'\\}
\end{center}
\begin{center}
	Bence Het\'enyi$^1$, Benjamin J. Brown$^{1,2}$, and Dominic J. Williamson$^3$ \\
    {\it \small $^1$IBM Quantum, IBM T. J. Watson Research Center, Yorktown Heights, NY 10598, USA\\
    $^2$IBM Denmark, Sundkrogsgade 11, 2100 Copenhagen, Denmark\\
    $^3$IBM Quantum, IBM Almaden Research Center, San Jose, CA 95120, USA
    }
    \\
	(Dated: \today)
\end{center}
\vspace*{0.1cm}

\onecolumngrid
\begin{center}
\parbox{400pt}{\small \hspace{5pt} Here we provide additional details on the construction and the numerical analysis.
In Sec.~\ref{sec:colorcode_hhx} we present different layouts to implement the color code on a heavy-hexagonal lattice. In Sec.~\ref{sec:magicstate_hhx}, using the same layout, we show how to implement the gauging logical measurement and provide explicit circuits to measure the Clifford stabilizers of the deformed code that emerges during the gauging measurement step. Sec.~\ref{sec:gauging_proof} reviews the construction of Ref.~\cite{williamsonLowoverheadFaulttolerantQuantum2024} in the context of logical Clifford measurements. Sec.~\ref{sec:opt} describes the stabilizer measurement and the flagged Clifford gauge measurement circuits on the square lattice. In Sec.~\ref{sec:comp_cult} we provide additional details on the comparison between our scheme and magic state cultivation \cite{chamberlandFaulttolerantMagicState2019,gidneyMagicStateCultivation2024}. Finally, the growing strategy of the cultivation stage is detailed in Sec.~\ref{sec:growing}, and strategies for transitioning to the surface code are also discussed.

}
\end{center}

\setcounter{equation}{0}
\setcounter{figure}{0}
\setcounter{table}{0}
\setcounter{section}{0}
\setcounter{page}{1}

\renewcommand{\theequation}{S\arabic{equation}}
\renewcommand{\thefigure}{S\arabic{figure}}
\renewcommand{\thesection}{S\arabic{section}}

\section{Color Code on the heavy-hex lattice}
\label{sec:colorcode_hhx}

In this appendix, we review the Color Code and the heavy-hex lattice. 
Below we depict a patch of the heavy-hex lattice. 
\begin{align}
\vcenter{\hbox{\includegraphics[page=1]{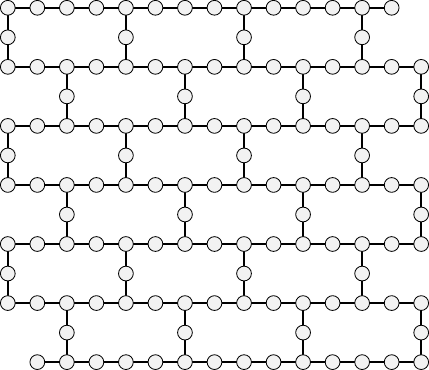}}} 
\end{align}
The vertex-edge lattice structure of this qubit layout is well suited to implement a gauging logical measurement of a transversal logical operator.  

It is convenient to introduce a 3-coloring of the plaquettes on the heavy-hex lattice when discussing the Color Code. 
This induces a 3-coloring of the edges on the heavy-hex lattice, shown below. 
\begin{align}
\vcenter{\hbox{\includegraphics[page=2]{Figures}}}
\label{fig:Floquet_lattice}
\end{align}
The plaquette stabilizers of the Color Code can be measured on the heavy-hev lattice in a variety of ways, including via the procedure proposed in Ref.~\cite{williamsonDynamicalQuantumCodes2025}. 

\subsection{Square patch}

We first discuss a square patch of the Color Code with green vertical, and red or blue horizontal, boundary conditions.  
With these boundary conditions, the Color Code encodes a pair of logical qubits that support a transversal Hadamard gate that implements a logical Hadamard+SWAP transformation. 
This Color Code also supports a logical $CZ$ gate that is implemented by a transversal gate given be a product of $\sqrt{-i}XS$ on \textit{up} vertices, depicted on the left below, and $\sqrt{i}XS^\dagger$ on \textit{down} vertices, depicted on the right below. For the remainder of this section we do not explicitly write the complex phase in front of $\sqrt{-i}XS$ and $\sqrt{i}XS^\dagger$, rather we write $XS$ and $XS^\dagger$ for notational simplicity, where the complex phases required to make these operators are left implicit. 
\begin{align}
\vcenter{\hbox{\includegraphics[page=10]{Figures}}} 
&&
\vcenter{\hbox{\includegraphics[page=11]{Figures}}} 
\label{eq:UpDownVertices}
\end{align}
There are three possible color configurations for the up vertices, we have depicted a single choice for simplicity, and similarly for the down vertices. 
The product of $XS$ and $XS^\dagger$ gates described above preserves the code space, as ${(XS) X (XS)^\dagger = Y}$, and ${(XS^\dagger) X (XS^\dagger)^\dagger = - Y}$. 
This transversal action maps the $X$-type plaquette operators to $Y$-type plaquette operators with the appropriate $(\pm 1)$ sign,  determined by the parity of half the number of vertices in the plaquette. 
Similarly, the logical action depends on the length of the logical representative up to a $(\pm 1)$ sign. 

Below, on the left we have green vertical and red horizontal boundaries. 
On the right we have green vertical and blue horizontal boundaries. 
\begin{align}
\label{CCSquare}
\vcenter{\hbox{\includegraphics[page=4]{Figures}}} 
\quad \vcenter{\hbox{\includegraphics[page=5]{Figures}}} 
\end{align}
Here the grey qubits are not required to implement the Color Code. 
We have included an extra corner qubit on both of the above Color Codes for convenience. 
This corner qubit, and the neighboring qubit, can be moved to nearby unused grey qubits to avoid adding extra physical qubits. 
Below we focus on the green and red boundary Color Code on the left.

\subsection{Triangle patch}

We now consider a triangular patch of the Color Code with a red, green, and blue, boundary. 
This configuration of the Color Code encodes a single logical qubit that supports transversal $S$ and $H$ gates which generate the full logical Clifford group. 
This includes the logical $XS$ gate, which has a transversal implementation as described above.
\begin{align}
\vcenter{\hbox{\includegraphics[page=21]{Figures}}} 
\qquad
\vcenter{\hbox{\includegraphics[page=22]{Figures}}} 
\label{eq:SteaneCode}
\end{align}
Again, the grey qubits are not required to implement these Color Codes. 
The smaller patch on the right is equivalent to Steane's code, which has distance three.

\section{Magic state preparation via gauging Clifford measurement on the Color Code} 
\label{sec:magicstate_hhx}

In this appendix, we describe the implementation of a gauging logical measurement of a transversal Clifford operator on the Color Code, which naturally obeys the heavy-hex connectivity constraints. 

We introduce a dynamical protocol to fault-tolerantly prepare a logical magic state in the Color Code using qubits laid out on the heavy-hex lattice with nearest neighbor gates and single qubit mid-circuit measurements with global classical communication and feedforward operations.
This can be used to prepare encoded magic states by initializing the Color Code in an appropriate logical state, such as $\ket{0}$, and measuring a logical Clifford operator, such as $\sqrt{i} X S^\dagger$. 
This code deformation creates a non-stabilizer code that is known to support nonabelian $D_4$ quantum double anyons~\cite{Prem2019,verresenEfficientlyPreparingSchrodingers2022}. 
Below, we further show that the deformed code is in fact the same as the one studied in Refs.~\cite{yoshida2015topological,Iqbal2023,davydovaUniversalFaultTolerant}. 
Interestingly, the non-Clifford logical operation is implemented by switching into and out of the nonabelian topological code and does not require creation or braiding of nonabelian anyons~\cite{davydovaUniversalFaultTolerant}.

To measure a transversal logical operator on the heavy-hex lattice we apply the following gauging measurement procedure~\cite{williamsonLowoverheadFaulttolerantQuantum2024}: 
\begin{enumerate}
    \item Perform single qubit rotations on all vertex qubits to bring the transversal logical operator on each site into the form of an $X$ operator.
    \item Initialize all edge qubits in the $\ket{0}$ state. 
    \item Perform CNOTs from each vertex to all surrounding edges. In each CNOT the vertex is the control qubit and the edge is the target qubit.
    \item Measure $X$ on each vertex qubit, and reset to $\ket{\pm}$ conditioned on the measurement outcome.
    \item Perform the CNOT circuit from step 3~again.
    \item Measure $Z$ on each edge qubit. 
    \item Apply a $Z$-type byproduct operator to vertex qubits. 
    \item Undo the single qubit rotations from step 1.
\end{enumerate} 
The above procedure measures the global transversal logical operator, but does not measure any other operator that is supported solely on the vertex qubits. 
The outcome of the logical measurement is recovered by taking the product of all vertex qubit $X$ measurement results. 
To implement the above procedure scalably and fault tolerantly requires $d$ rounds of quantum error correction for the deformed quantum code that appears between steps 4 and 5, where $d$ is the desired fault distance~\cite{williamsonLowoverheadFaulttolerantQuantum2024}. 
Alternatively, fault tolerance can be achieved without scalability by simply repeating the whole gauging measurement procedure as written a sufficient number of times and post-selecting on a consistent measurement outcome or taking a majority vote.

We remark that to perform steps 1 and 8 of the procedure for a Clifford operator requires the application of single site non-Clifford rotations from the third level of the Clifford hierarchy. 
Steps 2-4 result in the following measurement, shown below for a single vertex. 
\begin{align}
\vcenter{\hbox{\includegraphics[page=8]{Figures}}} 
\end{align}
The qubits initialized in $\ket{0}$ above are edge qubits adjacent a single vertex qubit, which is measured in the $X$-basis, denoted by the box labeled $X$. 
In the full procedure, each internal edge qubit is adjacent to a pair of vertices and consequently it is the target of two CNOT gates. 
Steps 5-6 result in the following measurement, shown below for a single edge. 
\begin{align}
\vcenter{\hbox{\includegraphics[page=3]{Figures}}} 
\end{align}
Here, the qubits initialized in $\ket{+}$ correspond to a pair of vertex qubits that are adjacent to an edge qubit which is measured in the $Z$-basis. 
More generally the vertex qubits are initialized in $\ket{\pm}$ depending on the outcome of the preceding vertex $X$ measurement in step 4. 
In the full procedure, each internal vertex qubit is adjacent to three edge qubits, and hence is the control for three CNOT gates. 
The byproduct operator in step 7 is found by taking a product of $Z_v$ operators on a collection of vertices whose $\mathbb{Z}_2$-coboundary ($\mathbb{Z}_2$-boundary on the dual lattice) matches the edge measurement results found during step 6, see Ref.~\cite{williamsonLowoverheadFaulttolerantQuantum2024} for an in depth description.

We now proceed to describe two examples. 
To provide a high level insight on these examples, we note that when the above gauging measurement procedure is applied to a global symmetry of a topological code, steps 1-4 implement a transformation on the anyon content of the code known as gauging a symmetry. 
Hence the deformed code that appears after step 4 supports anyons that correspond to gauging the global symmetry on the anyon theory of the initial code.  
There is a general formalism to calculate the effect that gauging has on a topological state, which depends on how the global symmetry acts on the initial code~\cite{barkeshli2014symmetry}. 
Below, we first consider a simple example where the global symmetry is the Color Code's $X$ logical operator, which acts on the anyon content of the theory in a trivial way. 
The result of gauging this symmetry on the anyon theory is simply to introduce another copy of the surface code. 
In the second example we consider gauging a global symmetry that permutes the anyon types of the Color Code in a nontrivial way by swapping their $X$ and $Y$ labels~\cite{Bombin2006}. 
In this case, it is known that gauging such a permutation symmetry on the abelian $\mathbb{Z}_2\times\mathbb{Z}_2$ anyons of the Color Code leads to a nonabelian $D_4$ anyon theory~\cite{barkeshli2014symmetry}. 

The second half of the gauging measurement procedure, steps 5-7, then correspond to \textit{ungauging} the global symmetry. 
This is equivalent to \text{condensing} the abelian anyons that correspond to gauge charges.  
In the first example, this simply condenses the $e$ anyons in one copy of the surface code and restores the anyon theory of the Color Code. 
In the second example, this condenses one of the abelian anyons in the $D_4$ theory to restore the $\mathbb{Z}_2\times \mathbb{Z}_2$ anyon theory of the Color Code.

There is a key distinction between our proposal and previous work on preparing nonabelian anyon states on a quantum device~\cite{Iqbal2023}. 
Our procedure demonstrates that by preparing a nonabelian $D_4$ state on a quantum device we can achieve a useful primitive for universal fault-tolerant quantum computation without the need to manipulate nonabelian anyons directly. 
This useful primitive is magic state preparation which can be combined with lattice surgery, or transversal, Clifford logic gates to perform universal fault-tolerant quantum computation in two dimensions without requiring the braiding of anyons or defects~\cite{davydovaUniversalFaultTolerant}. 

\subsection{Logical $X$ measurement}

First, we consider a simple example where no rotation is performed at step 1. 
In this case the transversal $X$ logical operator is measured directly on the Color Code. 
Steps 2-4 essentially initializes a new copy of surface code. 
The resulting deformed code at step 4~is equivalent to three copies of the surface code supported on the red, green, and blue colored edge sublattices, respectively. 
This deformed code is shown for the square patch Color Code from Eq.~\eqref{CCSquare} below. 
\begin{align}
\vcenter{\hbox{\includegraphics[page=7]{Figures}}} 
\end{align}
Similarly, for the triangular patch Color Code in Eq.~\eqref{eq:SteaneCode}, measuring the transversal $X$ logical operator leads to a deformed code equivalent to three copies of the surface code with boundary conditions shown below. 
\begin{align}
\vcenter{\hbox{\includegraphics[page=26]{Figures}}} 
\end{align}
Steps 5-7 then measure out the newly initialized surface code, and restore the original Color Code. 

\subsection{Logical $XS^\dagger$ measurement}

We now discuss applying the gauging measurement procedure to the transversal $\sqrt{-i}XS$, $\sqrt{i}XS^\dagger$, operator on Color Code. 
For the remainder of this section we do not explicitly write the phase factors that make the $XS,$ $XS^\dagger$, operators Hermitian. 
For a square patch of Color Code, see Eq.~\eqref{CCSquare}, this measures a logical $CZ$ gate. 
Similarly, for a triangle patch of Color Code, see Eq.~\eqref{eq:SteaneCode}, this measures a logical $XS^\dagger$ (or $XS$) gate. 
By initializing the square or triangle patch Color Codes in the $\ket{++}$ or $\ket{0}$ logical states, respectively, the above measurements produce encoded $CZ$ or $T$ ($T^\dagger$) magic states.

To perform the gauging measurement on the transversal $XS$, $XS^\dagger$, operator we first consider a symmetrized set of stabilizer checks for the code space where each check remains invariant under the symmetry action. 
In particular, for each plaquette $p$ we have a $Z$-type check, $\prod_{v\in p} Z_v$, and another check
\begin{align}
    B_p:={\frac{1}{2}(\prod_{v\in p} X_v \pm \prod_{v\in p} Y_v)=\frac{1}{2} (\mathds{1} + \prod_{v \in p} Z_v ) \prod_{v\in p} X_v } ,
\end{align}
where the $(\pm 1)$ sign is determined by the parity of up-type vertices in plaquette $p$. 
The $B_p$ operator satisfies ${B_p^2=\frac{1}{2} (\mathds{1} + \prod_{v \in p} Z_v )}$ and hence acts as 0 outside the $\prod_{v \in p} Z_v = +1$ subspace, while it acts as $\prod_{v\in p} X_v$ and hence has $(\pm 1)$ eigenvalues within the $\prod_{v \in p} Z_v = +1$  subspace. 

Step 1 of the gauging measurement procedure involves a $T$ rotation on each up-vertex and a $T^\dagger$ rotation on each down-vertex, see Eq.~\eqref{eq:UpDownVertices}. 
This leaves the $Z$-type plaquette checks invariant, and maps the $B_p$ terms to
\begin{align}
    B_p \mapsto B_p' &= \frac{1}{2^{4}}(\prod_{v\in p} (X_v -\varepsilon_v Y_v) - \prod_{v\in p} (Y_v +\varepsilon_v X_v)) 
    \\
    &= \frac{1}{2^{4}} \prod_{v\in p} X_v \, (\prod_{v\in p} (\mathds{1} -\varepsilon_v i Z_v) +  \prod_{v\in p} ( \mathds{1} + \varepsilon_v i Z_v )) 
    \\
    &= \frac{1}{2^{3}} \prod_{v\in p} X_v \, (
    \mathds{1} + \sum_{v<v'\in p} i^{\epsilon_v+\epsilon_{v'}} Z_v Z_{v'} + \sum_{v<v'<v''<v'''\in p} i^{\epsilon_v+\epsilon_{v'}+\epsilon_{v''}+\epsilon_{v'''}} Z_v Z_{v'} Z_{v''} Z_{v'''} + \prod_{v \in p} Z_v
    ) 
    \label{eq:RotatedBp}
    \\
    & = \frac{1}{2^{3}} \prod_{v\in p} X_v \, (
    \mathds{1} + \prod_{v\in p} Z_v
    )(
    \mathds{1} + \sum_{v,v'\in p} i^{\epsilon_v+\epsilon_{v'}} Z_v Z_{v'} 
    ) ,
    \label{eq:RotatedBp2}
\end{align}
where $\varepsilon_v=+1$ for up type vertices, and $\varepsilon_v=-1$ for down type vertices. 
In the sums over vertices above we use a vertex ordering starting at the bottom left corner of a plaquette and proceeding counterclockwise. 
The operator above is written for a hexagonal plaquette with six verties, for a boundary plaquette with four vertices we instead have 
\begin{align}
    B_p' &= \frac{1}{2^{2}} \prod_{v\in p} X_v \, (
    \mathds{1} + \sum_{v,v'\in p} i^{\epsilon_v+\epsilon_{v'}} Z_v Z_{v'} + \prod_{v \in p} Z_v
    ) .
\end{align}

After step 4 of the gauging measurement procedure the code has been deformed into a new code supported on the edge qubits. 
While the outcomes of the $X_v$ measurements at step 4 are random, we can apply a byproduct operator to arrive at the deformed code corresponding to all $X_v=+1$ outcomes. 
The byproduct operator is a product of $Z$ string operators on edges that end on the $X_v=-1$ outcome vertices and on open boundaries. 
For simplicity we focus on the $X_v=+1$ outcome deformed code.

The deformed code is stabilized by three plaquette checks per bulk plaquette, and two per boundary plaquette. 
Two of the bulk plaquette checks are $Z$-type, and take the form $Z_1Z_3Z_5,Z_2Z_4Z_6,$ using the numbering convention for the edges around a plaquette shown below. 
These $Z$-type plaquettes originate from deformed $Z$-type plaquette checks in the Color Code, combined with additional stabilizer checks that are introduced on the edge qubits initialized in $\ket{0}$. 
\begin{align}
\vcenter{\hbox{\includegraphics[page=9]{Figures}}} 
\label{eq:PlaqNumbering}
\end{align}
The $Z$ plaquette checks are products of $Z$ operators on edges of a single color in the boundary of the plaquette. 
A circuit to measure the $Z_1Z_3Z_5$ check is shown below, where the vertex and edge qubits in the boundary of the plaquette are ordered starting from the vertex qubit in the bottom right corner, and proceeding counterclockwise. 
\begin{align}
\label{eq:3ZPlaq}
\vcenter{\hbox{\includegraphics[page=12]{Figures}}} 
\end{align}
An alternative circuit with fewer measurements can also be used. 
\begin{align}
\label{eq:3ZPlaqb}
\vcenter{\hbox{\includegraphics[page=13]{Figures}}} 
\end{align}
Here, the long range CNOT gates must be mediated by swaps or bridge qubits~\cite{Hetenyi2024}. 
Similar circuits can be used to measure $Z_2 Z_4 Z_6$
The circuit in Eq.~\eqref{eq:3ZPlaq} for both $Z$-type plaquette checks can be implemented simultaneously using swaps as follows. 
\begin{align}
\vcenter{\hbox{\includegraphics[page=14]{Figures}}} 
\end{align}
A similar approach applies to simultaneously implement the circuit in Eq.~\eqref{eq:3ZPlaqb} for both $Z$-type plaquette checks. 
Boundary plaquettes have three edges, which are numbered similar to the convention above. 
\begin{align}
\vcenter{\hbox{\includegraphics[page=20]{Figures}}} 
\end{align}
The boundary $Z_1 Z_3$ check can be measured similar to above via the following circuit. 
\begin{align}
\vcenter{\hbox{\includegraphics[page=15]{Figures}}} 
\end{align}
In the $Z$-type check measurement circuits above we have reinitialized all physical qubits in the $\ket{0}$ state. 
This is to ensure that the post-measurement state lies in the $(+1)$-eigenspace of all $Z$-type plaquette checks, which is required to gain useful information from a subsequent measurement of the $\widetilde{B}_p$ check. 

The remaining plaquette check is derived by deforming the symmetrized plaquette check in the Color Code, see Eq.~\eqref{eq:RotatedBp}, 
\begin{align}
    \widetilde{B}_p  
    =& \frac{1}{2^{4}} \prod_{e \perp p} X_e \, (\mathds{1}+\prod_{e \in p} Z_e) 
    \nonumber \\
    & \quad (
    \mathds{1} + \sum_{v<v'\in p} i^{\epsilon_v+\epsilon_{v'}} Z_{\langle v,v' \rangle} + 
    \sum_{v<v'<v''<v'''\in p} i^{\epsilon_v+\epsilon_{v'}+\epsilon_{v''}+\epsilon_{v'''}} Z_{\langle v,v'\rangle }Z_{\langle v'',v'''\rangle} + Z_1 Z_3 Z_5
    ) ,
    \label{eq:GaugedBp}
\end{align}
where the notation $e \perp p$ denotes the edges that meet the boundary of plaquette $p$ at a single vertex, in Eq.~\eqref{eq:PlaqNumbering} these are the red edges.
The notation $\langle v, v' \rangle$ denotes the counterclockwise path from $v$ to $v'$ in the boundary of plaquette $p$, and $Z_{\langle v,v'\rangle}=\prod_{e \in \langle v,v'\rangle} Z_e$. 
For the term $Z_1 Z_3 Z_5$ we have used the edge numbering from Eq.~\eqref{eq:PlaqNumbering}. 
We remark that no $Z_6$ terms appear in the sum of $Z$ terms in the second line of Eq.~\eqref{eq:GaugedBp}, as they are factored out in the $(\mathds{1}+\prod_{e \in p} Z_e)$ term. 
We note that $(\mathds{1} + Z_1Z_3Z_5)$ can be further factored out from the sum of $Z$ terms in the second line of Eq.~\eqref{eq:GaugedBp}, which follows from deforming the $ (\mathds{1} + \prod_{v\in p} Z_v)$ term that is factored out in Eq.~\eqref{eq:RotatedBp2}. 
We then have 
\begin{align}
    \widetilde{B}_p  
    =& 
    \frac{1}{2^{4}} \prod_{e \perp p} X_e \, (\mathds{1} + Z_1Z_3Z_5) (\mathds{1} + Z_2Z_4Z_6)
    (
    \mathds{1} + Z_1 + Z_2 - Z_1Z_2 + Z_4 + Z_5 - Z_4Z_5 + Z_1 Z_4 
    \nonumber 
    \\
    & \qquad \qquad + Z_1 Z_5   - Z_1 Z_4 Z_5 + Z_2 Z_4 + Z_2 Z_5 - Z_2 Z_4 Z_5 - Z_1 Z_2 Z_4 - Z_1 Z_2 Z_5 + Z_1 Z_2 Z_4 Z_5
    ) 
    \\
    =&
     \prod_{e \perp  p} X_e \, \frac{1}{2}(\mathds{1} + Z_1Z_3Z_5) \frac{1}{2}(\mathds{1} + Z_2Z_4Z_6) \,
     CZ_{12} CZ_{45} \,
    ,
\end{align}
where we have used the factoring out of the $ Z_1Z_3Z_5$ and $Z_2Z_4Z_6$ terms to ensure that $Z_3$ and $Z_6$ do not appear in the additional sum over $Z$ terms in the first equation. 
We have used that $CZ_{12}=\frac{1}{2} ( \openone +Z_1 +Z_2 -Z_1Z_2)$ in the second equation. 
We remark that $\widetilde{B}_p$ can be written in a more symmetric fashion as 
\begin{align}
    \widetilde{B}_p = 
     \prod_{e \perp  p} X_e \, \frac{1}{2}(\mathds{1} + Z_1Z_3Z_5) \frac{1}{2}(\mathds{1} + Z_2Z_4Z_6) \,
     CZ_{12} CZ_{23} CZ_{34} CZ_{45} CZ_{56} CZ_{16} \,
    ,
\end{align}
using the fact that $ CZ_{23} CZ_{34} CZ_{56} CZ_{16}$ acts trivially in the $Z_1Z_3Z_5=+1,Z_2Z_4Z_6=+1,$ subspace. 
This demonstrates that the choice of starting point in the numbering of edge qubits involved in $\widetilde{B}_p$ is arbitrary. 
To measure $\widetilde{B}_p$ we can first measure $Z_1Z_3Z_5,Z_2Z_4Z_6,$ and apply a correction to ensure we are in the $Z_1Z_3Z_5=+1,Z_2Z_4Z_6=+1,$ subspace and then  measure the simpler operator 
\begin{align}
    \widetilde{B}_p' = \prod_{e \perp  p} X_e \, CZ_{12} CZ_{45} .
\end{align}
This approach means that the measured value of $\widetilde{B}_p'$ is only useful for inferring the measured value of $\widetilde{B}_p$ when we can ensure the measured values of the $Z$-type plaquette operators are $Z_1Z_3Z_5=+1,Z_2Z_4Z_6=+1$. 
This feature requires special care when decoding, and can be handled by a just-in-time or on-the-fly decoder~\cite{bombin20182dquantumcomputation3d,Bombin2018Transversal,Brown2020a,davydovaUniversalFaultTolerant}. 
A circuit on the heavy-hex lattice to measure $\widetilde{B}_p'$ is shown below. 
\begin{align}
\vcenter{\hbox{\includegraphics[page=16]{Figures}}} 
\end{align}
In the above we use the ordering of qubits following the numbering below. 
\begin{align}
\vcenter{\hbox{\includegraphics[page=17]{Figures}}} 
\end{align}
In the circuit above the unmeasured $\ket{0}$ and $\ket{+}$ states are required to be in the specified state to measure the desired $\widetilde{B}_p'$ operator. 
These states can be measured in the appropriate basis to provide additional flag information to detect errors. 
An alternative circuit with fewer CNOT gates can be found by removing the final two CNOT gates between qubits 8 and 9 and measuring qubit 9 in the $X$ basis, in this case the measurement result is the product of the $X$ measurement results on qubits 7 and 9. 

Similar to the bulk deformed plaquette operator, the deformed boundary plaquette check can be written 
\begin{align}
    \widetilde{B}_p  = \prod_{e \perp  p} X_e \, \frac{1}{2}(\mathds{1} + Z_1Z_3)\,  CZ_{12} ,
\end{align}
where the number of edges that meet the boundary plaquette at a single vertex can be two, three, or four, depending on the boundary condition. 
Similar to above, in the $Z_1 Z_3=+1$ subspace we can measure the simpler operator
$\widetilde{B}_p' = \prod_{e \perp  p} X_e \, CZ_{12}$. 
A circuit to implement this measurement on the heavy-hex lattice is shown below. 
\begin{align}
\vcenter{\hbox{\includegraphics[page=18]{Figures}}} 
\end{align}
Above, the qubits are ordered as follows. 
\begin{align}
\vcenter{\hbox{\includegraphics[page=19]{Figures}}} 
\end{align}

In the appendices of Ref.~\cite{Prem2019} the deformed code described above, up to a change of basis on individual qubits, was demonstrated to be equivalent to the nonabelian $D_4$ quantum double anyon theory. 
Here, we have further shown that the deformed code is in fact the same as the one studied in Refs.~\cite{yoshida2015topological,Iqbal2023,davydovaUniversalFaultTolerant}, up to a codespace-preserving redefinition of the checks. 

The smallest nontrivial example of the $XS$ measurement can be performed on a distance three triangle patch of Color Code, shown embedded into the heavy-hex lattice in Eq.~\eqref{eq:SteaneCode}. 
In this case the deformed code is defined on the edges of the patch shown below. 
\begin{align}
\vcenter{\hbox{\includegraphics[page=23]{Figures}}} 
\label{eq:SmallPatchNumbers}
\end{align}
In this case the stabilizer group of the deformed code commutes as it is generated by the checks
\begin{align}
    \langle Z_1 Z_4,\ Z_2 Z_5,\  Z_3 Z_6,\ X_1 X_4 CZ_{23},\ X_2 X_5 CZ_{13},\ X_3 X_6 CZ_{12} \rangle ,
\end{align}
where we have used the numbering convention in Eq.~\eqref{eq:SmallPatchNumbers}. 
This code is equivalent to three Bell pairs on qubits $14,25,36,$ that have been entangled via a $CCZ$ gate on qubits $123$. 
This code can only detect errors due to a reduction to the Color Code distance induced by the code deformation. 
However, the correction chosen to return to the code space does not matter for the $XS$ logical measurement in this example, and on larger triangle patches, as the code space contains only the desired logical magic state. 

A larger example of the $XS$ measurement can also be implemented on heavy-hex using the embedding a distance five triangular Color Code shown on the left of Eq.~\eqref{eq:SteaneCode}. 
This results in a deformed code that is similar to three copies of the surface code on the lattices shown below, in that it has the same $Z$-type plaquette terms. 
However the $X$-type star terms in the deformed code involve additional $CZ$ operators, which are indicated via colored triangles that connect each vertex to pairs of edge qubits below. 
\begin{align}
    \vcenter{\hbox{\includegraphics[page=24]{Figures}}} 
\end{align}
The three sublattices that appear are depicted separately below, for clarity. 
\begin{align}
    \vcenter{\hbox{\includegraphics[page=25]{Figures}}} 
\end{align}

\section{Gauging logical measurement proof}

\subsection{Logical action}
\label{sec:gauging_proof}

In this appendix, we prove that the gauging logical measurement scheme of Ref.~\cite{williamsonLowoverheadFaulttolerantQuantum2024} works without assuming that the logical operator is a Pauli product. Let us take a logical operator of the product form $C_L=\prod_i U_i$, where $U_i = U_i^{-1}$ and $i$ runs over the data qubits. Further, let us define an ancilla system where each ancilla is connected to a small subset of data qubits. If every ancilla is connected to a pair of qubits, one can define gauge operators for every data qubit
\begin{equation}
A_i = U_i\prod_{a\in \partial i}X_a. 
\end{equation}
The product of these operators is the logical operator $C_L$ we aim to measure. The claim is that one can measure these gauge operators and {\it (i)} the product of measurements $\sigma_L=\prod_im_i$ is the logical eigenvalue and {\it (ii)} the measurement outcome of the ancilla qubits will be sufficient to restore the correct post-measurement state $(1+\sigma_L C_L)\ket{\psi}_Q$.

Taking an arbitrary code state, and initializing the ancillas in the $\ket 0$ state upon measuring the gauge operators individually we get
\begin{eqnarray}
    \prod_{Q'\subset Q,\left|Q^\prime\right|\le\frac{\left|Q\right|}{2}}\left(\prod_{i\in Q'}{m_iU_i}+\prod_{i\in Q/Q'}{m_iU_i}\right)\ket{\psi}_Q\otimes\ket{v_{Q'}}_A\,,
\end{eqnarray}
where $m_i$ are the measurement outcome of gauge $A_i$, the summation goes through every $Q'$ subset of data qubits, and $\ket{v_{Q'}}_A$ is a product state on the ancillas where each ancilla is $\ket 0$ ($\ket 1$) if it is connected to an even (odd) number of data qubits in $Q'$. Exploiting that $U_i = U_i^{-1}$ we can rewrite the equation above as
\begin{eqnarray}
    (1+\sigma C_L)\prod_{Q'\subset Q,\left|Q^\prime\right|\le\frac{\left|Q\right|}{2}}\prod_{i\in Q'}{m_iU_i}\ket{\psi}_Q\otimes\ket{v_{Q'}}_A\,.
\end{eqnarray}
Once the ancillas are measured in a certain state $\ket{v_{Q'}}_A$ the state (up to a global phase) becomes
\begin{eqnarray}
    (1+\sigma C_L)\prod_{i\in Q'}{U_i}\ket{\psi}_Q\,,
    \label{smeq:postmeas_state}
\end{eqnarray}
and the correct post-measurement state can be restored if the measured state $\ket{v_{Q'}}_A$ determines the subset $Q'$ modulo $Q$. 

We define the ancilla graph $G$ by identifying data qubits with vertices and ancilla qubits with the edges of the graph. In that case, the subset $Q'$ of data qubits can be determined by taking an arbitrary edge path from an arbitrarily chosen qubit $j$ to every other qubit $i\neq j$, and include $i$ in the subset only if the parity of ancilla measurement outcomes along the edge path is odd. 

The strategy for constructing a suitable ancilla graph and determining the correction operator from the measurement can be found in the main text.

\subsection{The role of the Cheeger constant}
\label{sec:cheeger}

\begin{figure*}
    \centering
\includegraphics[width=.35\linewidth]{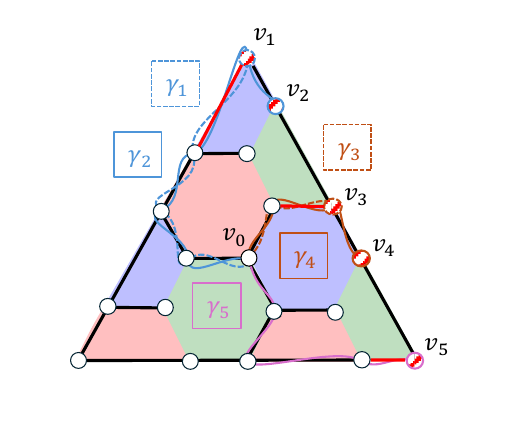}
    \caption{Role of the Cheeger constant. Data qubit nodes, together with black edges form an ancilla graph $G$ with $h(G)\approx1/2$ on the $d=5$ color code. Three edges of $G$ are colored red, indicating measurement errors. Every wiggly path $\gamma_i$ that is connected to one of qubits on the right edge $v_i$, must cross an odd number of faulty ancillas leading to five faulty Clifford corrections (indicated by white/red pattern).}
    \label{fig:Cheeger}
\end{figure*}

Here we discuss the role of the Cheeger constant of the ancilla graph $h(G)$ in the case of the Gauging Clifford measurement. In Ref.~\cite{williamsonLowoverheadFaulttolerantQuantum2024} the authors prove that the $h(G)\geq 1$ is sufficient for the gauged code to have the same code distance as the original code. 

Although in this work we do not measure the stabilizers of the gauged code, $h(G)\geq 1$ still provides a sufficient condition for the distance to be preserved. To show this, we write down the correction operator from Eq.~\eqref{smeq:postmeas_state} explicitly as
\begin{equation}
    \mathcal C(\{m_a\}_{a\in A}) = \prod_{i=1}^n U_{v_i}^{[\sum_{a\in\gamma^*_i}m_a]},
\end{equation}
where $\gamma^*_i\subset A$ is a subset of ancillas along an arbitrary edge path between data qubit $v_0$ and qubit $v_i$ and $m_a\in\{0,1\}$ is the measurement result on ancilla $a$ after the gauge measurements. Consequently the correction will be wrong if there was a measurement error along the path. 

Cycles in the ancilla graph protect against individual measurement errors, but some sets of measurement errors separate the graph into two subgraphs, such that any path connecting the two subgraphs must cross a faulty edge (ancilla) an odd number of times. Consequently, if the designated node ($v_0$) is on one subgraph, faulty correction will be applied on every qubit of the other subgraph. The Cheeger constant being smaller than one implies that $t$ measurement errors could induce $t/h(G)>t$ data qubit errors.

Fig.~\ref{fig:Cheeger} shows an example of a $d=5$ color code with a connected graph $G$ that has three cycles and a Cheeger constant of $h(G)<1$. Five data qubits along the support of a logical operator can receive faulty corrections by having three measurement errors on the edges connecting them to the rest of the graph.

\section{Optimizing the circuit depth of the gauging measurement}
\label{sec:opt}

\begin{figure*}
    \centering
    \includegraphics[width=1\linewidth]{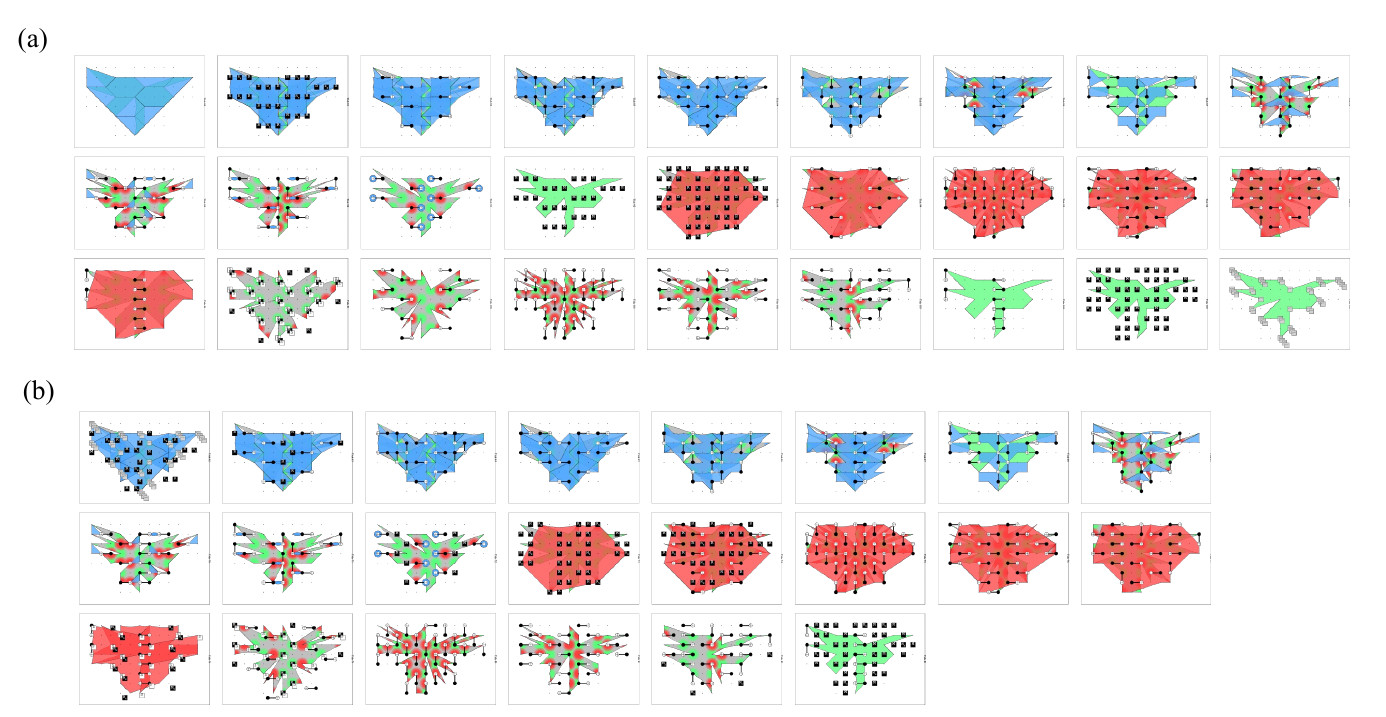}
    \caption{(a) The same circuit as in the main text. (b) Pipelined circuit with 18 CX time steps and 3 measurement/reset time steps.}
    \label{fig:pipelining}
\end{figure*}

To measure the color code stabilizers we use the superdense syndrome measurement circuit of Ref.~\cite{gidney_new_2023} and prepare the ancilla Bell pair on the two ancillas that are inside each plaquette through the flag qubits. 
This can be done by initializing the ancillas and the flag qubit and applying three CX gates, i.e., $CX_{a_2,f}CX_{f,a_2}CX_{f,a_1}\ket{0}_{a_1}\ket{+}_f\ket{0}_{a_2}$. 
Since every data qubit on a given plaquette is the nearest neighbor of either $a_1$ or $a_2$, we can follow Ref.~\cite{gidney_new_2023} to encode the X and Z plaquette information into the Bell state of the ancillas. We alternate the role of the ancillas between neighboring columns of hexagons, and shift the superdense CX schedule by two time steps between the left and right sides of each plaquette. The measurement of the Bell state is performed reversing the role of the ancillas. This allows us to shorten the circuit depth by one (10 CX layers instead of 12). See the frst 13 panels of Fig.~\ref{fig:pipelining}(a) for the contracting Z-type (expanding X-type) detecting regions.

In the main text we argued that a single flag qubit can be used to flag every edge in the honeycomb lattice. Before discussing the order in which the CX gates of the gauge measurements and flag qubits are applied we make a short comment on a possible optimization step we did not fully explore in this work. From the point of view of the minimum-weight logical, flagging the edge parallel to the boundary is not necessary. However, we find that omitting the boundary-parallel edges from the flagging for every color does reduce the fault distance, which is believed to be a consequence of the weight-reduction of higher-weight logicals. Therefore, in this work, we included every edge in the flagging.

The CX scheduling of the flagged gauging measurement step is based on the following logic. Here we describe the schedule from the perspective of the three ancillas that belong to a plaquette, instead of the measurement of a single gauge operator. After resetting every ancilla and the flags we perform five CX layers. In the first one a CX is targeted on one of the three ancillas; in the second, the first edge-ancilla is flagged and another one is targeted from a data qubit; in the third step the first edge is finished, the second is being flagged and the third one is targeted from a data qubit; the fourth step finishes the second edge, flags the third, before the third edge would be finished in the last step. We then proceed by measuring the data qubits. Afterwards, the CX layers are repeated, which completes the gauge measurements, and the ancillas are measured along with the flags. Finally, based on the ancilla outcomes, we restore the codespace using physical Clifford operations. See the last 14 panels of Fig.~\ref{fig:pipelining}(a) where the contracting detector compares the logical Y measurement of this round with that of the previous round, and the expanding detector compares this measurement with the next one.

In order to reduce the number of idling operations in the circuit, each of which comes with a single-qubit depolarizing channel, we performed some circuit optimizations that preserve the effective CX ordering of the superdense circuit and the flagging of the gauging measurement. Fig.~\ref{fig:pipelining} shows the circuits from the main text as well as the pipelined version where we allow measurements to be performed at the same timestep as CX gates.

\section{Growing and transitioning to a matchable code}
\label{sec:growing}

\begin{figure*}
    \centering
    \includegraphics[width=1.\linewidth]{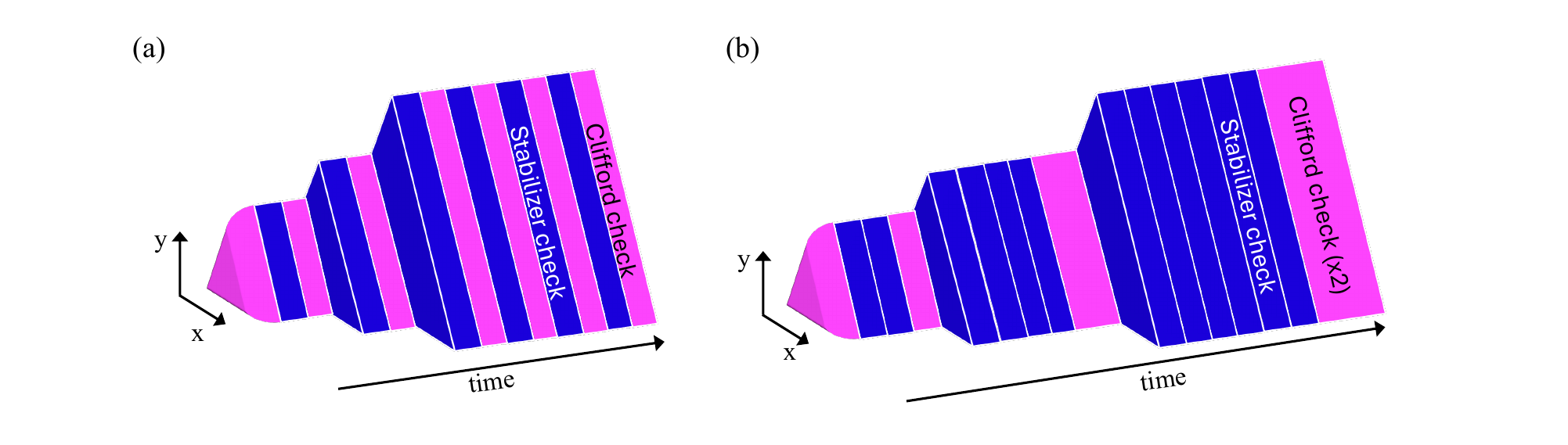}
    \caption{(a) Gauging Clifford measurements on a growing the lattice (intermediate code distances $d_\text{int}\in\{3,5,7\}$). Starting with a unitary injection, one stabilizer measurement and one Clifford check is performed at every stage until the target distance, where the remaining $(d+1)/2$ rounds take place. (b) Cultivation in the strictly distance-preserving scheme with $d-1$ stabilizer rounds before every Clifford double check. Blue slices denote stabilizer measurements and imply growing with physical Bell-state preparations unless preceded with a Clifford check of the same distance.}
    \label{fig:growing}
\end{figure*}

To prepare magic states with larger fault distance $d_f$ we start from the $d=3$ example and prepare magic state non fault-tolerantly using a unitary circuit. This is followed by a stabilizer measurement and a gauging Clifford check at $d_\text{int}=3$. We grow the lattice by preparing Bell states on the edges that match the color of the boundary we are growing along. We continue this until we reach the final target distance, where we perform an additional $(d+1)/2$ rounds of stabilizer and Clifford check pairs. See Fig.~\ref{fig:growing}(a) for a process diagram of the growing stages. The reason we cannot do two Clifford checks at every intermediate distance is the existence of spacetime-diagonal errors, that start at the previous spatial boundary and terminate at the opposite corner of the final code in last time step. Such an error flips the logical state at the last growing step, and mutes every logical detector along the way. Such errors are the consequence of the growing step, and have nothing to do with the gauging measurement scheme. In fact, these type of errors are the reason why cultivation requires $d-1$ stabilizer rounds before the Clifford double-check at every intermediate distance (see Fig.~\ref{fig:growing}(b))\cite{gidneyMagicStateCultivation2024}.

\subsection{Number of stabilizer rounds in cultivation}
\label{sec:cultivation_depth}

In this subsection, we argue that the number of stabilizer rounds required at each intermediate distance in \cite{gidneyMagicStateCultivation2024} is $d-1$ when higher-distance color codes are considered. This is a consequence of the error-propagation properties of the superdense syndrome circuit, which we detail below.

\begin{figure}
    \centering
    \includegraphics[width=0.65\linewidth]{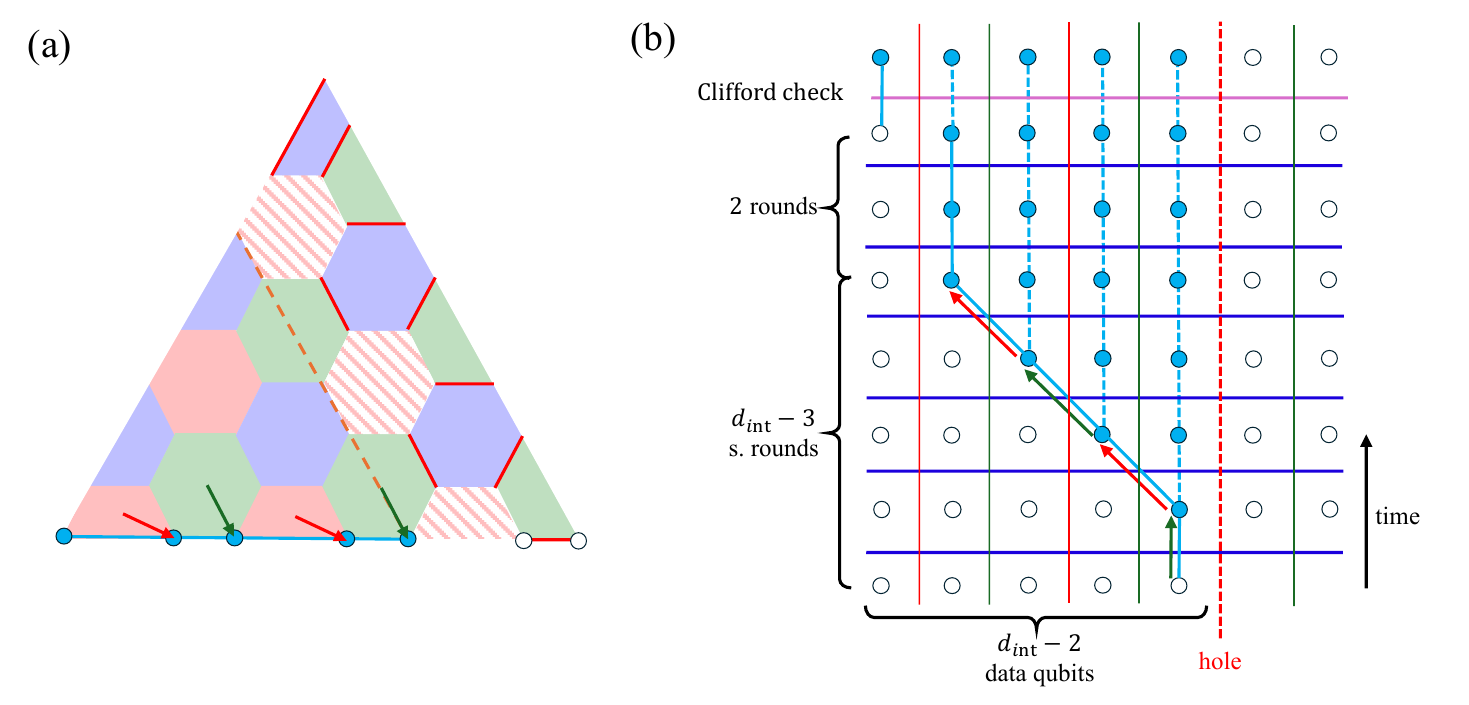}
    \caption{(a) A low-weight logical Z error (cyan data qubits) in the $d=7$ color code at the growing stage ($d=5\to7$) and hook errors (colored arrows) that can inject a Z error along the logical without being detected in the same round. Dashed plaquettes are not deterministic at the growing stage. (b) Spacetime picture of the undetectable logical error, showing the diagonal hooks and the subsequent measurement errors. Vertical lines denote the stabilizer that two neighboring qubits share, while the dashed red stabilizer (hole) is not deterministic. }
    \label{fig:spacetime_hooks}
\end{figure}

After the growing step, the stabilizers that have the same color as the extended boundary are not deterministic. Therefore, an error string can terminate on them right after the growing (see Fig.~\ref{fig:spacetime_hooks}(a)). The error string of length $d_\text{int}-2$ could terminate at the corner of the same type. However, if such error happened in the same round, the Clifford check would catch it. 

The only way for such an error to go undetected is to have an even number of data errors before the Clifford check and an odd number afterwards that makes the string undetectable for the final perfect syndrome round. In a phenomenological noise setting, such $d_\text{int}-2$ error strings can be eliminated by having two stabilizer rounds before the Clifford check, which would require two more readout errors to occur before an undetectable failure, increasing the distance to the desired $d_\text{int}$.

Hook errors in the superdense syndrome extraction circuit can inject Z errors in certain locations without being detected in the same round (see Fig.~\ref{fig:spacetime_hooks}(a)). I.e., Z errors that propagate from the ancilla qubits during the Z stabilizer mapping stage, and the support on the ancilla are removed during the X mapping. Since the X stabilizer mappings happen after the Z error has propagated to the data qubit, a string of Z errors can propagate round by round towards a boundary without being detected in any of the rounds (see Fig.~\ref{fig:spacetime_hooks}b). This way, an even-weight error string arrives at the Clifford check stage, and one last data error can complete the logical Z error which can not be detected by any number of stabilizer rounds afterward.

Similar arguments can be made to show that $d_\text{int}-1$ rounds are also necessary at every intermediate stage. Instead of going through every possibility, we simulated the $d=7$ cultivation protocol with every configuration, i.e., between $1$ and $d_\text{int}-1$ rounds at every intermediate stage $d_\text{int}$. The results in Tab.~\ref{tab:distance} show that $d_\text{int}-1$ is indeed the minimal choice that achieves the fault distance $7$ under circuit level noise.

\begin{table}[h]
\centering
\begin{tabular}{ccc|cc}
\hline
\hline
  \multicolumn{3}{c|}{number of syndrome rounds} & \multicolumn{2}{c}{resulting distance}
  \\
  \hline
  $d_\mathrm{int}=3$ & $d_\mathrm{int}=5$ & $d_\mathrm{int}=7$ & $d_\mathrm{graph}$ & $d_\mathrm{circ}^*$
  \\
\hline
1 & 1 & 1 & 5 & 4 \\
1 & 1 & 2 & 6 & 4 \\
1 & 1 & 3 & 6 & 5 \\
1 & 1 & 4 & 6 & 5 \\
1 & 1 & 5 & 6 & 5 \\
1 & 1 & 6 & 6 & 5 \\
1 & 2 & 1 & 5 & 4 \\
1 & 2 & 2 & 6 & 5 \\
1 & 2 & 3 & 7 & 5 \\
1 & 2 & 4 & 7 & 5 \\
1 & 2 & 5 & 7 & 5 \\
1 & 2 & 6 & 7 & 5 \\
1 & 3 & 1 & 5 & 5 \\
1 & 3 & 2 & 6 & 5 \\
1 & 3 & 3 & 7 & 5 \\
1 & 3 & 4 & 7 & 5 \\
1 & 3 & 5 & 7 & 6 \\
1 & 3 & 6 & 7 & 6 \\
1 & 4 & 1 & 5 & 5 \\
1 & 4 & 2 & 6 & 5 \\
1 & 4 & 3 & 7 & 5 \\
1 & 4 & 4 & 7 & 5 \\
1 & 4 & 5 & 7 & 6 \\
1 & 4 & 6 & 7 & 7 \\
\hline
\hline
\end{tabular}
\caption{%
  Fault distance of the $d=7$ magic-state cultivation circuit as a function of
  the number of stabilizer-measurement rounds inserted at each intermediate
  code distance.
  The cultivation sequence is
  unitary injection;
  syndrome rounds and Clifford check at $d_\mathrm{int}=3$;
  growing $(3{\to}5)$;
  syndrome rounds and Clifford check at $d_\mathrm{int}=5$;
  growing $(5{\to}7)$;
  syndrome rounds and Clifford check at $d_\mathrm{int}=7$.  $d_\mathrm{graph}$ is the graphlike fault distance from \texttt{stim}'s
  \texttt{shortest\_graphlike\_error}.
  $d_\mathrm{circ}^*$ is the weight of the shortest undetectable error from \texttt{search\_for\_undetectable\_logical\_errors}. The target distance is $d=7$. Note that only one round of stabilizer measurement is required at $d_\text{int} = 3$ (instead of $d_\text{int}-1$) because the unitary injection ensures deterministic stabilizer eigenvalues in the first step.
}
\label{tab:distance}
\end{table}

\subsection{Escape stage}

\begin{figure*}[h]
    \centering
    \includegraphics[width=1\linewidth]{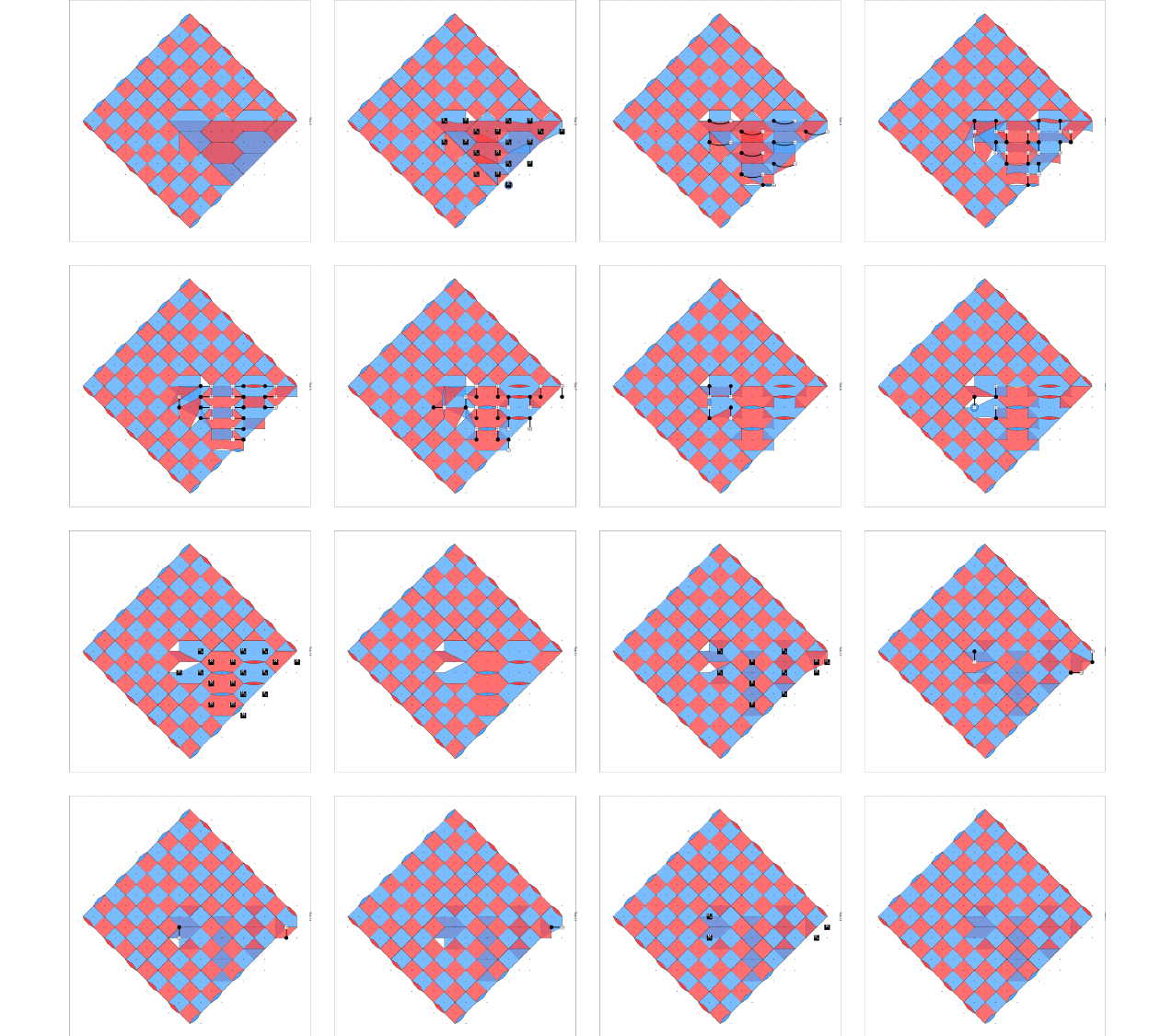}
    \caption{The full circuit for the transition of the grafted surface code to a regular (rotated) surface code. These steps cover the first three diagrams in Fig.~\ref{fig:simulations}(c).}
    \label{fig:escape_full}
\end{figure*}

Furthermore, in the escape stage~\cite{gidneyMagicStateCultivation2024} it is possible to deform the color code portion into a regular surface code patch, as shown in Fig.~\ref{fig:simulations}(c). This opportunity arises from the commensurability of our color code lattice with the surface code. Including the flag qubits (step 11 on Fig.~\ref{fig:escape_full}) in the deformed color code allows one to keep most of the original detectors in the color code region. The regular surface code patch has a shallower syndrome measurement circuit and hence potentially better performance than the matchable grafted code of Ref.~\cite{gidneyMagicStateCultivation2024}. Fig.~\ref{fig:escape_full} shows a circuit that converts the grafted surface code into a regular surface code. The last round of the transformation (not shown) is a regular surface code stabilizer measurement cycle, where the appropriate products of measurement outcomes (involving at most two) are included in the corresponding detectors of the last step shown in Fig.~\ref{fig:escape_full}. These deformation circuits were found using \texttt{crumble} \cite{gidney_stim_2021}.

The conversion circuit above has been analyzed in two different settings. The first option is to convert the grafted code into a surface code right away (before measuring the grafted stabilizers $d$ times). The advantage of this would be that the stabilization stage can be done in a matchable code without additional post-selection on color-code detectors. Unfortunately, this approach performs very poorly under circuit level noise because some weight-two errors are undetectable where the two codes connect. The second approach, where the grafted code is stabilized before transitioning to the surface code, has little advantage over the scheme proposed by Ref.~\cite{gidneyMagicStateCultivation2024}, since the color code region does not contribute to the code distance. This case is equivalent to discarding the grafted part of the matchable code.

In conclusion, the escape stage would need to the prescription of Ref.~\cite{gidneyMagicStateCultivation2024}, or rely on a post-selected lattice surgery scheme similar to that shown in Ref.~\cite{hiranoEfficientMagicState2025}.

\section{Comparison to cultivation}
\label{sec:comp_cult}

\begin{table*}[t]
    \centering
    \begin{tabular}{ccc}
        \hline
         & Cultivation & This work \\        
        \hline
        single Clifford check time & $(3d_\text{int}-1)/2\, \CXtime + 2\, \MRtime$ & $8\,\CXtime + 3\,\MRtime$ \\
        stabilizer time & $8\,\CXtime + 2\,\MRtime$ & $10\,\CXtime + 2\,\MRtime$\\
        \# stabilizer rounds & $(d^2-1)/4$ & d-1\\
        total time $/\, (d-1)$ & $(11d+15)/4\, \CXtime + (d+5)/2\,\MRtime$ & $18\, \CXtime + 5\,\MRtime$\\
        \hline
    \end{tabular}
    \caption{Comparison of the circuit depth between cultivation and the gauging Clifford measurement protocol. The comparison assumes a final color code path of distance $d$ and the patch is gradually grown achieving $d_f = d$. In the table, $3\leq d_\text{int} \leq d$ is the code distance of the color code patch in the intermediate stage, $\MRtime$ is the measurement and reset duration, and $\CXtime$ is the duration of the CX gates. The number of Clifford checks is $d-1$ in both cases, but cultivation requires $d_\text{int}-1$ rounds after every growing step, and performs a double-check at every intermediate distance as shown in Fig~\ref{fig:growing}.}
    \label{tab:circuit_depth}
\end{table*}

In order to determine the break-even point in success probability between the gauging Clifford measurement and cultivation, we summarize the time overhead of both protocols in Tab.~\ref{tab:circuit_depth}. Note that in the table the time overhead of the Clifford check in the cultivation scheme is divided by two to account for the fact that the Clifford check circuit increases the time distance by two. This is in fact the biggest practical advantage of cultivation over the gauging measurement scheme. The gauging Clifford measurement scheme, on the other hand, brings two important improvements {\it (i)} the Clifford check has a constant-depth circuit {\it (ii)} the protocol requires $d-1$ stabilizer and Clifford check rounds in total as opposed to cultivation, where each growing step requires $d-1$ stabilizer rounds (see Fig.~\ref{fig:growing}(b) for illustration). Taking stock, the break-even point in total circuit depth should be $d=5$, however, this is not the most relevant figure of merit from a practical standpoint.

Fig.~\ref{fig:simulations}(b) shows the success probability of the gauging Clifford measurement vs. cultivation employing the respective growing strategies from Fig.~\ref{fig:growing}.
Under the uniform circuit-level noise model used in both works, the breakeven point in success probability only comes as $d=9$. This is due to the qubit overhead of the gauging protocol. The latter uses $50\%$ more qubits, resulting in similar number of fault locations only at $d=9$~\footnote{For this rough estimate, we assume that the success rate depends the same way on the number of fault locations in both cases, and the number of fault locations is the total number of qubits times the total time. Further, we take $t_\text{CX}=t_\text{M/R}$ which describes well the uniform noise model.}. 
We note that the most important metric would be the success probability at a given logical error rate, but Monte-Carlo sampling for the logical error rate becomes unfeasible already at $d=7$.

Finally, we note that the fault-distance of the cultivation scheme with a $d=5$ color code in Ref.~\cite{gidneyMagicStateCultivation2024} is only $d_f = 4$. While weight-4 errors do not limit the performance of the protocol at $p_\text{phys} = 0.1\%$ due to the low multiplicity of such errors, at lower error rates the $d_f = 5$ gauging protocol is expected to have an advantage.

\clearpage

\end{document}